\documentclass[%
 reprint,
 superscriptaddress,
 amsmath,amssymb,
 aps,
]{revtex4-2}

\usepackage{graphicx}% Include figure files
\usepackage{dcolumn}% Align table columns on decimal point
\usepackage{bm}% bold math

\usepackage{hyperref}% add hypertext capabilities
\begin{document}

\preprint{APS/123-QED}

\title{Structural aging of a cohesive and amorphous \\ granular solid under cyclic loading}% Force 

\author{William Hobson-Rhoades}
\affiliation{%
 University of Michigan\\
 Department of Mechanical Engineering
}%
\author{Douglas J Durian}
\affiliation{
 University of Pennsylvania\\
 Department of Physics and Astronomy% with \\
}%
\author{Yue Fan}
\affiliation{%
 University of Michigan\\
 Department of Mechanical Engineering
}%
\author{Hongyi Xiao}
 \email{hongyix@umich.edu}
\affiliation{%
 University of Michigan\\
 Department of Mechanical Engineering
}%

\date{\today}% It is always \today, today,

\begin{abstract}
We investigate how cyclic loading evolves the structure and deformation behaviors of a granular raft composed of particles floating at an air-oil interface.
The raft has a disordered particle packing structure, and is cohesive due to capillary interactions between particles.
Under uniaxial cyclic loading with a small strain amplitude, the raft's packing structure experiences an aging process characterized by logarithmically increasing packing fraction and decreasing structural heterogeneity. 
The observed structural change is due to particle dynamics that are organized around morphologically evolving voids in the raft.
The raft is then subjected to quasi-static tension or compression tests until failure. In comparison with non-aged rafts, the rafts that experienced cyclic loading show a higher strength, higher stiffness, and lower ductility, along with qualitatively different features, such as a stress overshoot in the loading curve. 
\end{abstract}


%\tableofcontents

\maketitle

\section{Introduction}
 
The deformation of amorphous solids is an important process throughout numerous industries, geophysical processes, and everyday life.\cite{bonn2017yield,nicolas2018deformation,berthier2025yielding}
%that our world would be unrecognizable without it. 
Processes like squeezing toothpaste, spreading mayonnaise, and scrubbing a foaming soap exemplify ductile deformation, where a material's constituent particles continuously rearrange without system-spanning fracture. 
Conversely, shattering glass, breaking chalk, and large-scale landslides exemplify brittle failure, characterized by the formation of shear bands and a steep drop in the loading curve.\cite{ozawa2018random,popovic2018elastoplastic,barlow2020ductile,singh2020brittle,yeh2020glass,priezjev2022mechanical,parmar2019strain,leishangthem2017yielding,pollard2022yielding,sharma2025activity}

Tuning the ductility of amorphous solids is important for many engineering applications. 
A common approach is thermal annealing, which uses sequential temperature changes to manipulate a material's structure and its associated energy landscape. Past studies have uncovered the roles of the annealing temperature, duration, and quenching rate on the ductility of various materials.\cite{yeh2020glass,das2022annealing,barlow2020ductile,priezjev2018molecular,utz2000atomistic,fan2017effects,shen2007plasticity,bouchbinder2013cooling,ketkaew2018mechanical}
As the interplay between structure, elasticity, and plasticity is complex in amorphous solids,\cite{nicolas2018deformation,zhang2022structuro,xiao2023identifying} a change in ductility is often accompanied by concurrent changes in other deformation parameters, such as the elastic modulus and yield strength.\cite{ozawa2018random,popovic2018elastoplastic,barlow2020ductile,singh2020brittle,yeh2020glass,priezjev2021shear,pollard2022yielding,sharma2025activity} 
However, this complex interplay also enables the manipulation of a disordered structure through mechanical deformation. 

An amorphous solid can be mechanically annealed toward lower energy states via cyclic shear.\cite{das2022annealing} The lower energy serves to enhance the material stability and generally decrease the ductility.\cite{yeh2020glass,ozawa2018random,popovic2018elastoplastic,das2022annealing,barlow2020ductile}
Numerical simulations have identified several important factors that influence the behaviors of amorphous solids during cyclic shear.\cite{keim2014mechanical,leishangthem2017yielding,priezjev2022mechanical,das2022annealing,jana2023relaxation,li2024infinitely} 
For example, large shearing amplitudes may induce shear bands,\cite{priezjev2021shear,priezjev2022mechanical} while small amplitudes can lead the particle system to a localized energy minimum, characterized by reversible particle trajectories.\cite{jin2018stability,keim2014mechanical,keim2022mechanical,slotterback2012onset,royer2015precisely}

However, studies on this topic often rely on simulations of idealized particle systems, while the complexity stemming from realistic particle interactions and alternative loading modes\cite{jana2020structural} has not been addressed.

While experiments measuring the change of bulk behaviors of amorphous solids such as metallic glasses have been informative,\cite{zhang2014mechanical,wang2015mechanical} obtaining the necessary particle-scale information for a mechanistic understanding is challenging. 
This renders experiments with macroscopic particles useful, such as granular particles and colloids, as the particle packing structure and dynamics during cyclic loading can be directly measured.\cite{mueggenburg2005behavior,richard2005slow,zhang2010statistical,slotterback2012onset,farhadi2015stress,xiao2022probing,zhao2022ultrastable,yuan2024creep} Existing experimental work with macroscopic particles has already revealed novel features in cyclic deformation, such as a perpetual creeping motion, potentially resulting from complexities in the energy landscape.\cite{xiao2022probing,yuan2024creep} However, most existing experiments on this topic used particles with steric repulsive interactions, which fail to consider more complicated effects such as long-range attractions and repulsion, and possible many-body effects that are generally associated with molecular and atomic solids.\cite{li2024infinitely}

To gain a more general understanding of the structural evolution of an amorphous solid during cyclic loading, we use a model experimental system that consists of a disordered and two-dimensional particle raft floating at an air-liquid interface.\cite{xiao2020softmatter,xiao2023identifying,to2023rifts,druecke2023collapse,protiere2023particle}
The particle interaction includes a contact-based short range repulsion, friction, and a long-range capillary attraction resulting from buoyancy and surface tension, which has a functional form close to that of atomic interactions.\cite{nicolson1949interaction} The attraction may also have many-body effects if the liquid surface distortion due to the presence of many nearby particles cannot be superposed.\cite{dalbe2011aggregation} In our previous study, we configured a uniaxial tension test with the cohesive raft having two free boundaries.\cite{xiao2020softmatter,xiao2023identifying} This offers a loading mode that contains both volumetric and deviatoric strains, and permits particle packing density change.

In this work, we measure the dynamics and structural evolution in a cohesive granular raft
subjected to cyclic uniaxial loading protocols.
Different from previous cyclic shear protocols that are volume-preserving, 
we observe a logarithmic increase in the particle packing fraction and a decrease in the structural heterogeneity in the raft, which resembles long-term aging in thermal systems.
We connect these overall structural changes to microscopic particle dynamics that are organized around void collapsing behavior during cyclic loading.
Finally, we discuss how the structural changes in the cyclic loading impact the mechanical behaviors of the raft when it is subjected to subsequent tension or compression deformation until failure.

\section{Experimental methods}
\label{Sec2}

\begin{figure}[t!]
\centering
  \includegraphics[width=0.45\textwidth]{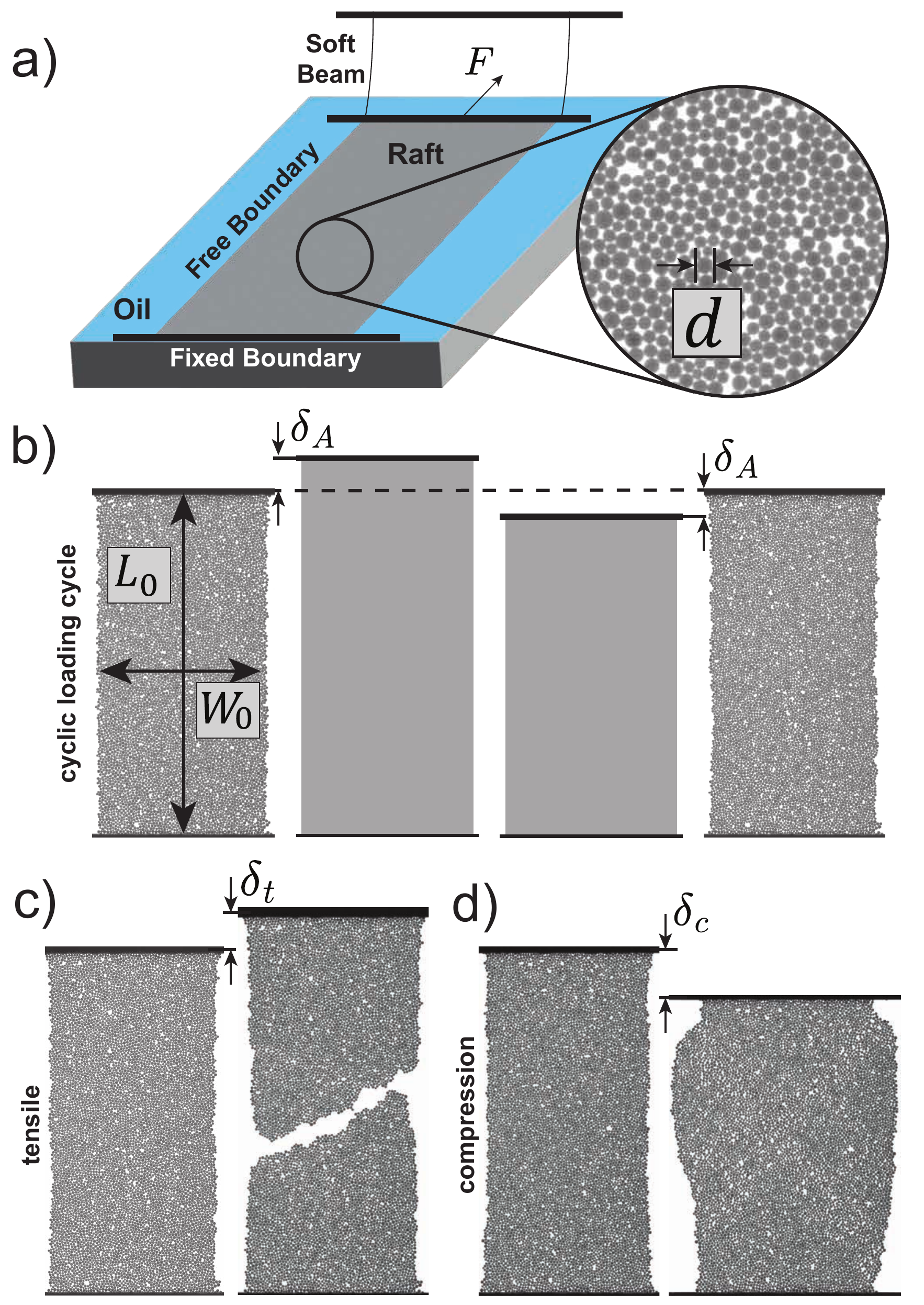}
  \vspace{-1 mm}
  \caption{a) Schematic of the experimental setup with an image of the particle packing. b) A schematic demonstrating the procedure for a single loading cycle. c) Images demonstrating a tensile loading test. d) Images demonstrating a compression loading test.}
  \label{fgr:capillary}
\end{figure}

Using a uniaxial deformation apparatus,\cite{rieser2015deformation,harrington2018anisotropic,harrington2020stagnant,xiao2020softmatter} we tested a granular raft composed of a monolayer of polydisperse spherical particles floating at an air-liquid interface, illustrated in \autoref{fgr:capillary}a. The particles are made of closed-cell Styrofoam, with a density of approximately 15~kg/m$^3$ and diameters of $d=1.0\pm$0.1~mm.\cite{xiao2020softmatter}
The liquid is a mineral oil, which has a density of $\rho=870\pm10$~kg/m$^3$ and a kinematic viscosity of $\nu=$13.5~cSt. Its surface tension is $\gamma=27.4\pm0.7$~dyn/cm, corresponding to a capillary length of $l_c=\sqrt{\gamma/\rho g}=1.8\pm0.2$~mm. This sets a small bond number of $Bo=d^2/4l_c^2\approx0.08$, at which $l_c$ approximates the interaction range of the capillary attraction between nearby particles.\cite{nicolson1949interaction,dalbe2011aggregation}

The granular raft has a rectangular shape with height $L_0=120d$ and width $W_0=60d$, corresponding to roughly 7000 particles. The two shorter sides are bonded to floating boundaries made of hollow carbon fiber tubes, visualized in \autoref{fgr:capillary}a. To generate the initial packing, we first manually laid a thin column of particles connecting the two floating boundaries. Then we sprinkled more particles near this column, prompting the particles to join the columnar assembly via capillary attraction. The generated rafts are dense and strictly monolayer. 

For the loading experiments, we fixed the position of the bottom boundary and used two soft beams to drive the top boundary with controlled displacement $\delta$. We measured the force, $F$, exerted on the top boundary by the particles by monitoring the small deflection of the beams, as described in our previous work.\cite{xiao2020softmatter}
The speed of the moving boundary is $\dot\delta=2\,\mu\mathrm{m}/\mathrm{s}$, setting a capillary number, $Ca=\nu\rho\dot\delta/\gamma\approx1\times10^{-6}$, meaning that the viscous drag on the particles is negligible in comparison to the surface tension-induced particle attraction. The strain rate under this speed is $\dot\delta/L_0=1.7\times10^{-5}$\,/s. We can therefore assume that the induced deformation is quasi-static.

Two types of experiments were conducted. The first is a uniaxial small amplitude cyclic deformation experiment. Each deformation cycle begins at $\delta=0$, after which the raft is extended to $\delta=\delta_A$. It is then compressed until $\delta=-\delta_A$, before finally returning to $\delta=0$, as shown in \autoref{fgr:capillary}b. Because performing thousands of loading cycles can be time consuming, we increased the driving speed to $\dot\delta=20\,\mu\mathrm{m}/\mathrm{s}$. 
A cyclic strain amplitude of $\epsilon_A=\delta_A/L_0=$0.3\% is selected, which will be discussed in \autoref{Sec3}.

In the second experiment, the raft is subjected to uniaxial tension ($\delta_t>0$) or compression ($\delta_c<0$) until it fails, as illustrated in \autoref{fgr:capillary}c and d, respectively.
To understand the influence of cyclic loading on the mechanical behavior of the raft, tension and compression tests were performed on the rafts directly following construction, as well as the ``aged'' rafts that had experienced cyclic loading up to cycle number $N=1000$. In addition to monitoring the force $F$, we imaged and tracked the trajectories of all particles in the system, while only using particles that are at least five layers away from the boundaries for analysis. Particle centers are tracked using an image analysis algorithm with sub-pixel accuracy.\cite{rieser2015deformation,xiao2020softmatter} 

\section{Effect of strain amplitude and number of cycles}
\label{Sec3}

\begin{figure}[h!]
\centering
  \includegraphics[width=0.48\textwidth]{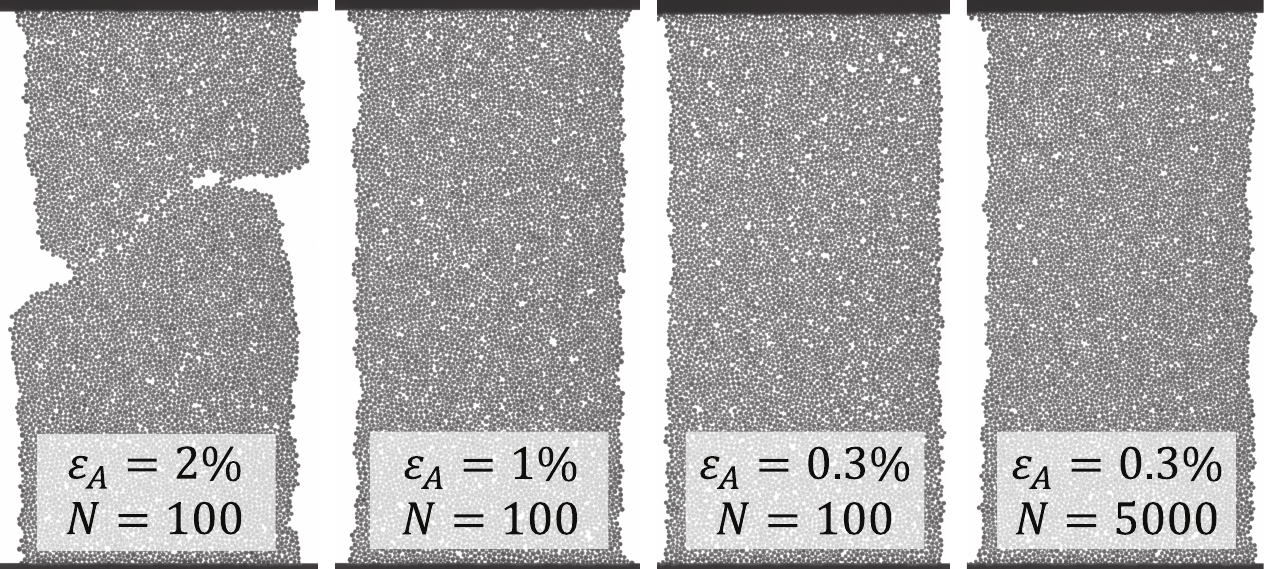}
  \vspace{-3 mm}
  \caption{Snapshots of granular rafts subjected to a variety of loading amplitudes at several different loading cycle numbers.}
  \label{fgr:ampcyc}
\end{figure}

Our preliminary results suggest that for $\epsilon_A > 1\%$, cyclic loading rapidly leads to the formation of a system-spanning shear band, which develops into a fracture at $N\approx100$, see an experimental snapshot in \autoref{fgr:ampcyc}. For $\epsilon_A \leq 1\%$, however, no such shear band forms, allowing the structure of the raft to evolve without global failure. The global strain of $1\%$ roughly coincides with the yield strain of the rafts during tensile tests as reported in our previous study\cite{xiao2020softmatter} and in \autoref{Sec7} of this study, placing our choice of $\epsilon_A=0.3\%$ in the quasi-elastic regime. This choice is further strengthened by another preliminary experiment in which a cyclic loading amplitude of $\epsilon_A=0.3\%$ was performed up to $N=5000$, where no drastic changes or signs of failure were observed in the structure of the raft. The comparison between $N=100$ and $N=5000$ for $\epsilon_A=0.3\%$ is also shown in \autoref{fgr:ampcyc}. 

\begin{figure*}[t!]
\centering
  \includegraphics[width=0.95\textwidth]{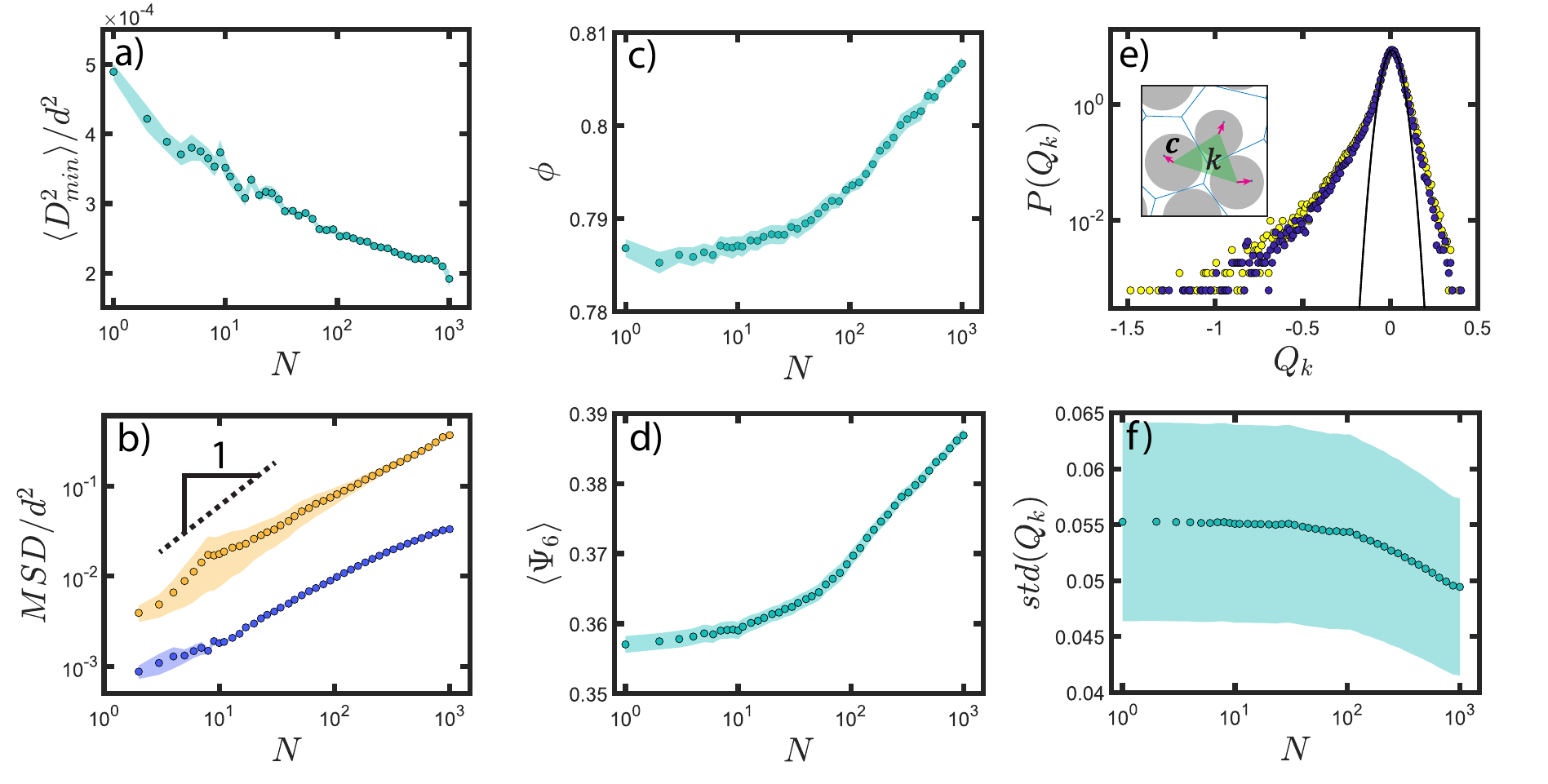}
  \vspace{-0.3 mm}
  \caption{Dynamic and structural quantities during cyclic loading with amplitude $\delta_A/L_0=0.03$ over the course of $N=1000$ cycles. a) Normalized particle mean-squared displacement vs. the cycle number. The orange curve is the original MSD and the blue is the cage-relative MSD. Slope of one provided for reference. b) The system-averaged $D^2_{\mathrm{min}}$ vs. the cycle number.
  c) The global packing fraction vs. the cycle number. d) The system-averaged bond order parameter $\Psi_6$ vs. the cycle number. e) Distribution of $Q_k$ for $N=0$ (yellow) and $N=1000$ (blue). The solid curve is a fit to the Gaussian distribution. Inset demonstrates the Voronoi cell anisotropy vectors for a single Delaunay triangle. f) The standard deviation of $Q_k$ vs. the cycle number. In all panels, the data points and shades respectively represent the mean and the test-to-test fluctuations from 19 repetitions.}

  \label{fgr:raft_struc}
\end{figure*} 

\section{Global dynamics during cyclic loading}
\label{sec4}

To quantify the degree of structural change between consecutive cycles, we use the non-affine displacement, $D^2_{\mathrm{min}}$, calculated for each particle $i$ by subtracting a best affine fit of its surrounding displacement field from the actual displacements of its neighbors $j$.\cite{falk1998dynamics} The formula is 
\begin{equation}
\label{eq:d2min}
D^2_{\mathrm{min},i}(t,\Delta t) = \frac{1}{n_j}\sum_{j=1}^{n_j} |\mathbf{r}_{ji}(t + \Delta t) - \mathbf{E}\mathbf{r}_{ji}(t)|^2,
\end{equation}
with $\mathbf{r}_{ji}=\mathbf{r}_{j}-\mathbf{r}_{i}$ as the relative position vector between particles $i$ and $j$, $n_j$ is the number of neighbors within 1.25$d$ of $i$, and $\mathbf{E}$ is a best affine transformation matrix.\cite{falk1998dynamics,li2015deformation} 
We consider the time interval $\Delta t$ to be one cycle and time $t$ to be the beginning of each cycle. 
 
The system and ensemble average, $\langle D^2_{\mathrm{min}}\rangle$, decreases nearly logarithmically as a function of the loading cycle $N$, see \autoref{fgr:raft_struc}a, indicating a slow-down in dynamics.

To quantify longer-term particle dynamics, we calculate the mean-squared displacement (MSD), $\langle \Delta r_i^2(t) \rangle = \langle |\mathbf{r}_i(t)-\mathbf{r}_i(0)|^2 \rangle$, with respect to $N=0$. As shown in \autoref{fgr:raft_struc}b, the MSD has two regimes: a super-diffusive regime with a slope above 1 for $N<10$ and a sub-diffusive regime with a slope below 1 for $N\ge10$. 

To better understand the observed super-diffusivity, we calculate a cage-relative MSD by subtracting the mean displacement of a particle's neighborhood from the displacement of that particle, written as 

\begin{equation}
\langle \Delta r_i^2(t) \rangle = \langle |\mathbf{r}_i(t)-\mathbf{r}_i(0) - \frac{1}{n_j}\sum_{j=1}^{n_j}[\mathbf{r}_j(t) - \mathbf{r}_j(0)]|^2 \rangle.
\label{eq:msd}
\end{equation}

\noindent This eliminates the effect of rigid body motion and leaves only the local affine and non-affine displacements. 
The result in \autoref{fgr:raft_struc}b shows a consistent sub-diffusive behavior for all shear cycles.
For $N<10$, the super-diffusivity in the original MSD is due to the rigid body motion of groups of particles. These groups may reside near locations of intense particle rearrangements, possibly originating from residual stresses due to the initial raft construction.

As previously noted, the structural evolution of our raft distinguishes itself from those with isochoric cyclic shearing simulations, as the presence of the free boundaries permits density change.

The density is tracked through the packing fraction $\phi$ calculated based on a Delaunay triangulation of the particle centers. For each triangle $k$, we calculate its area $A_k$ and the sum of the circular sector areas, $A^k_\mathrm{sc}$, due to the overlap area between $k$ and its three constituent particles (highlighted in green in \autoref{fgr:void_calc}b).

The packing fraction is then calculated as $\phi = \sum_k A_\mathrm{sc}^k / \sum_k A_k$. 

Over the course of 1000 loading cycles, \autoref{fgr:raft_struc}c shows that $\phi$ consistently increases, with a clear long-term logarithmic trend emerging for $N > 100$.
 
To detect higher-order structural changes accompanying the densification, we characterize local ordering through the widely used bond orientation parameter for each particle $i$, 

\begin{equation}
\Psi_{6,i} = \frac{1}{n_j}|\sum_{j=1} ^{n_j} e^{6i\theta_j}|,
\label{eq:psi6}
\end{equation}
with $\theta_j$ being the angle between the vector $\mathbf{r}_{ji}$ and a chosen reference direction, and $n_j$ in this case denotes the first shell of neighbors from the triangulation. 
 
Under this definition, a hexagonally ordered arrangement of neighbors around particle $i$ results in $\Psi_{6,i}\approx1$, and it decreases with increasing disorder.
Similar to the packing fraction, $\langle \Psi_6\rangle$ gradually increases, with a logarithmic trend for $N > 100$, as shown in \autoref{fgr:raft_struc}d, indicating increased ordering in the raft. 
 
We then characterize the packing structure's spatial fluctuations based on the Delaunay triangulation and the corresponding Voronoi tessellation. Specifically, we calculate the weighted divergence of the Voronoi cell anisotropy vector,\cite{rieser2016divergence,xiao2020softmatter} $Q_k$, which can be calculated for each Delaunay triangle $k$ as
\begin{equation}
\label{eq:qk}
Q_k \equiv (\nabla \cdot \mathbf{c})(A_k/\langle{A}\rangle),
\end{equation}
where $\mathbf{c}$ is the Voronoi cell anisotropy vector pointing from a particle's center to the centroid of its Voronoi cell, as illustrated in the inset of \autoref{fgr:raft_struc}e. The divergence in triangle $k$ is evaluated based on the $\mathbf{c}$ vectors of the three constituent particles,\cite{rieser2016divergence} which is further weighted by the ratio between the triangle area $A_k$ and the average triangle area $\langle A \rangle$ to ensure a zero mean. Under this definition, a triangle covering a tightly packed region should have its $\mathbf{c}$ vectors pointing outward, resulting in $Q_k>0$, and a triangle covering a loosely packed region should have $\mathbf{c}$ pointing inward and $Q_k<0$.

The distributions of $Q_k$ for $N=0$ and $N=1000$ are shown in \autoref{fgr:raft_struc}e, and both are Gaussian-like around $Q_k = 0$. For larger $|Q_k|$, the distributions have tails that trend above a Gaussian fit, signaling the existence of regions with highly heterogeneous local packing fractions in the raft. 
When comparing the distributions of $Q_k$ from $N=0$ and $N=1000$, a decrease in the standard deviation, $std(Q_k)$, is observed for larger values of $N$, which, similar to $\phi$ and $\Psi_6$, behaves logarithmically for $N>100$ as in \autoref{fgr:raft_struc}f. 

To help contextualize our results thus far, we can draw an analogy between the mechanically driven raft and a thermal system. The amplitude of the excitation, $\epsilon_A$, was held constant, which is analogous to a constant temperature. Additionally, the long-term logarithmic structural evolution after $N\approx100$ is analogous to the aging of a thermal glass below the glass transition temperature, which evolves toward lower energy states characterized by densification, more homogeneous structure, and increased ordering.
We therefore refer to this process as mechanical aging. In terms of dynamics, the decrease in non-affine motion (\autoref{fgr:raft_struc}a) is similar to previous molecular dynamic simulation results,\cite{priezjev2022mechanical} which suggest that particles should evolve toward configurations with lower potential energies.
The slow and long-term increase in the cage-relative MSD (\autoref{fgr:raft_struc}b) is also reminiscent of the MSD observed in aging thermal glasses.\cite{seoane2018spin,hammond2020experimental}
We note that similar correspondences between mechanical driving and thermal activation have been recognized in other disordered solids, particularly metallic glasses. Atomistic studies have demonstrated that external stress can markedly accelerate relaxation dynamics - effectively mimicking an increase in temperature - through facilitated exploration of a complex, fractal-like energy landscape.\cite{cao2017understanding,liu2021emergent} Moreover, stress–temperature scaling relations have been proposed in steady-state flow,\cite{guan2010stress} indicating the existence of quantitative mappings between mechanical and thermal variables. These precedents suggest that the resemblance uncovered here is not incidental but may reflect a more general principle of glassy dynamics. In this light, the present experimental system could, in the future, provide a controlled platform to probe and possibly quantify the correspondence between mechanical forcing and effective thermal aging in granular glasses. A systematic exploration of such mappings, however, is beyond the scope of this manuscript and warrants further study.

%\HX{-----------------------}

\section{Void-driven microscopic dynamics}
\label{sec:five}

\begin{figure}[h!]
\centering
  \includegraphics[width=0.45\textwidth]{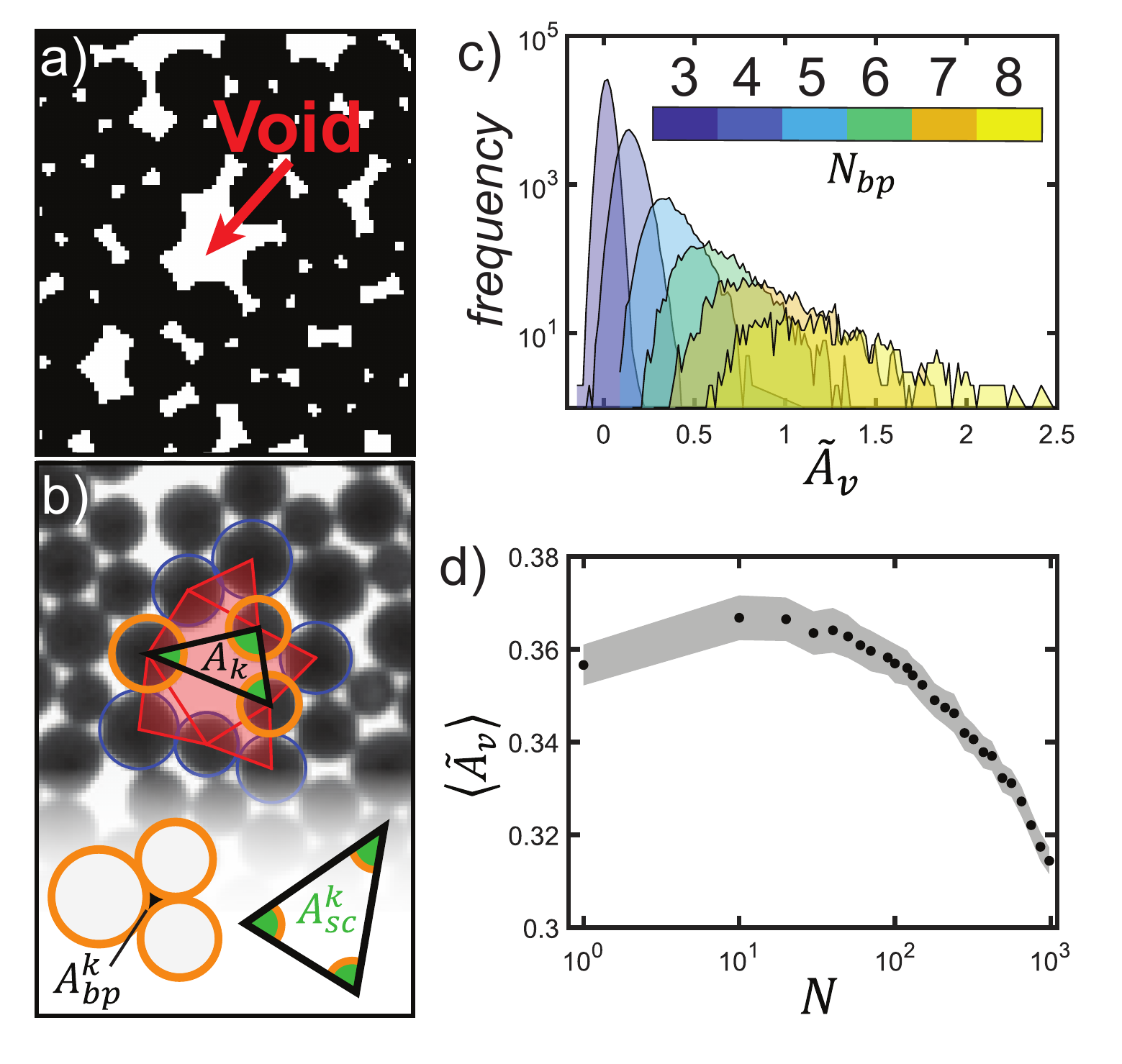}
  \vspace{-3 mm}
  \caption{Void identification and void area calculation. 
  %(a) An experimental image overlayed with Delaunay triangles. 
  a) A binarized image showcasing an identified void.
  %, after removing voids composed of three particles in close contact. 
  b) The original experimental image corresponding to a), overlaid with its Delaunay triangles highlighted in red. An example triangle $k$ is highlighted with black borders, and the associated particles are highlighted in orange. The circular sectors of the particles are highlighted in green. The insets illustrate the areas $A^k_{\mathrm{bp}}$, $A^k_{\mathrm{sc}}$, and $A_k$ that are calculated for each triangle $k$.
  c) Distributions of void areas based on the number of constituent particles making up the boundary of a void across all experiments taken at $N=1000$. d) A plot of the system-averaged void size across all experiments against the cycle number. The shaded region indicates standard error across repeated experiments.
  }
  \label{fgr:void_calc}
\end{figure}

The increase in packing fraction and decrease in structural heterogeneity observed during the aging process point towards the possibility that loosely packed regions are especially prone to rearrangements that result in local densification and spatial homogenization.
To quantify this possible interplay between structure and deformation, we identify voids from experimental images by
%We determine the local density of a region by characterizing voids, which represent the space between particles as discrete regions. 
%To identify voids, we begin by 
distinguishing the particle phase from the background oil phase through binarization, with an example shown in \autoref{fgr:void_calc}a. To evaluate the area of a void, we again rely on the Delaunay triangulation, which partitions each void into triangles, where the vertices of the Delaunay triangles fall on the boundary particles surrounding a void.  For each triangle $k$, we calculate three areas, including the triangle area $A_k$, the sum of the particle circular sector areas, $A^k_\mathrm{sc}$, and the enclosed area, $A^k_\mathrm{bp}$, as the area between three constituent particles if they are closely packed. All area types are depicted in \autoref{fgr:void_calc}b. The void area $A_v$ is then calculated as
\begin{equation}
\label{eq:void}
A_v = \sum_{k=1}^{n_\textrm{vt}} A_k-A^k_\mathrm{sc}-A^k_\mathrm{bp},
\end{equation}
where $n_\textrm{vt}$ denotes the number of constituent triangles for each void. Under this definition, a void surrounded by three closely packed particles (as in \autoref{fgr:void_calc}b) will have zero area, as there is no room for its area to further reduce. 

In \autoref{fgr:void_calc}c, we show the distributions of normalized void areas, $\tilde{A}_v=4A_v/\pi d^2$, which are categorized by the number of their constituent particles $N_{\textrm{bp}}$. For voids with larger $N_{\textrm{bp}}$, both the average and the variance of the void areas are larger.

Note that although voids with $N_{\textrm{bp}}=3$ should have $\tilde{A}_v=0$, some non-zero values exist due to measurement noise, and they should be neglected as they are only small fractions of a particle's projected area. 

We then track the system-averaged void areas across all repeated experiments as $N$ increases. Initially, $\langle\tilde{A}_v\rangle$ increases slightly, before decreasing nearly logarithmically after $N=100$ to a value significantly lower than the initial average void size. This behavior is consistent with the packing fraction curve in \autoref{fgr:raft_struc}c, where there is an initial decrease in the packing fraction, followed by a logarithmic increase after $N=100$. The plot of system-averaged void area against $N$ can be seen in \autoref{fgr:void_calc}d.

\begin{figure}[t!]
\centering
  \includegraphics[width=0.48\textwidth]{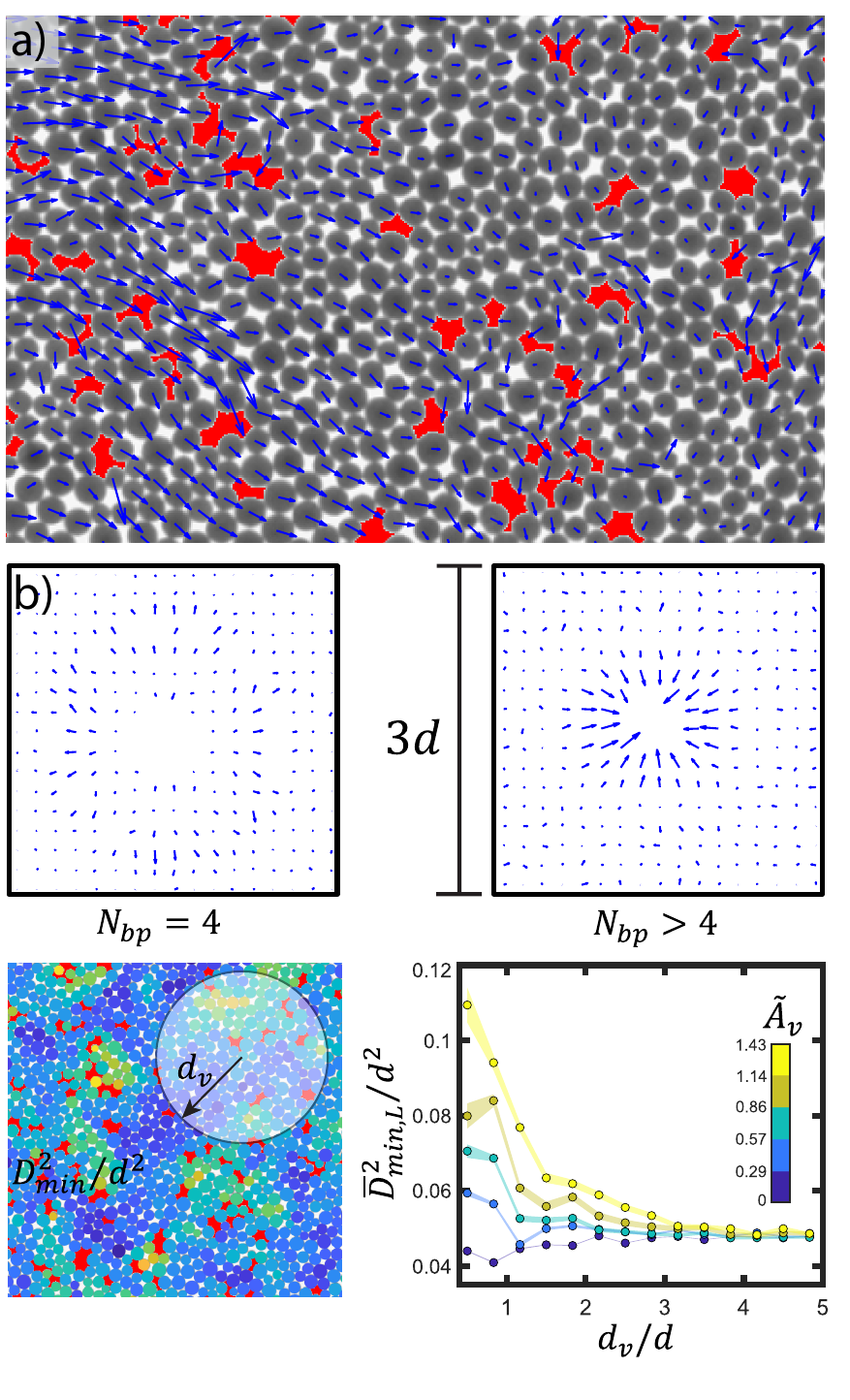}
  \vspace{-8 mm}
  \caption{Dynamics of particles near voids. a) An experimental image overlaid with particle trajectories (blue) and large identified voids (red). Trajectories are from $N=0$ to $N=1000$. b) Relative displacement field around  a void from $N=0$ to $N=1000$ with $N_\textrm{bp}=4$, averaged across all experiments. c) Average displacement field around a void from $N=0$ to $N=1000$ with $N_\textrm{bp}>4$ across all experiments. d) Particles colored by $D^2_{\mathrm{min},L}$ with large voids highlighted in red. e) The averaged $\bar{D}^2_{\mathrm{min},L}$ of particles as a function of their distance to the closest void, plotted for different void sizes. 
  }
  \label{fgr:void_dyn}
\end{figure}

To understand how voids influence the dynamics of the raft during structural aging, we examine the particle displacement field of the raft, with an example shown in \autoref{fgr:void_dyn}a. Specifically, we compute the average displacement field around each void by spatially binning the particle displacement as a function of the particle position relative to the centroid of each void and subtracting the average background displacement. We then discount all bins with less than 20 data points.

The resulting displacement field, when calculated for $N_\textrm{bp} =4$, only exhibits small radially outward displacements roughly one particle diameter away from the center, as shown in \autoref{fgr:void_dyn}b. When $N_\textrm{bp} >4$, however, a noticeable tendency for particles to move toward the void's center is observed, as shown in \autoref{fgr:void_dyn}c. A slightly higher displacement magnitude is observed in the direction of the free boundaries, highlighting the effect of their presence. This result suggests that the displacement field of the raft may be primarily driven by voids with $N_\textrm{bp} >4$, and that voids with $N_\textrm{bp}=4$ may have little impact on the dynamics of the raft.

To further quantify the dynamics of the raft near voids, we examine the non-affine particle motion close to the identified voids, with an example visualized in \autoref{fgr:void_dyn}d. As a void typically evolves over many loading cycles, we evaluate \autoref{eq:d2min} using particle positions from the initial and the final cycle of an experiment, yielding $D^2_{\mathrm{min},L}$, with the subscript $L$ introduced to distinguish from the frame-by-frame $D^2_{\mathrm{min}}$. For each void, we establish a local reference at the void's centroid and average the $D^2_{\mathrm{min},L}$ of the nearby particles at different distances, $d_v$. The resulting average, $\bar{D}^2_{\mathrm{min},L}(d_v)$, shown in \autoref{fgr:void_dyn}e, indicates that the non-affinity generally rises close to voids and is higher for larger voids.
 %It is also observed that non-affinity close to a void increases with area, suggesting that the largest individual changes to structure come from the largest voids.
All void sizes seem to have the same radius of influence of $d_v/d\approx3$, at which $\bar{D}^2_{\mathrm{min},L}$ converges to the background value.
Voids with $\tilde{A}_v<0.29$, however, tend to have slightly lower $\bar{D}^2_{\mathrm{min},L}$ when measured close to the void than the background value measured far away. Based on \autoref{fgr:void_calc}c, the area threshold of $\tilde{A}_v<=0.29$ roughly corresponds to $N_\textrm{bp} \le 4$, and higher bins are mostly populated by voids with $N_\textrm{bp} > 4$. It follows that voids with areas above the $\tilde{A}_v=0.29$ threshold contribute significantly more to the change in structure of the raft, which may be due to lower rearrangement energy barriers in close proximity to them. If so, we may expect that a particle located in a region with a high density of voids may be particularly susceptible to rearrangements that result in densification, some instances of which can be seen in \autoref{fgr:void_dyn}a and \autoref{fgr:void_dyn}d.

\section{Evolution of void morphology}
\label{sec:void}
 
Although the dynamics of particles near voids have been explored, we have yet to investigate the way that a void's morphology changes as a result of those dynamics. Therefore, we further quantify the evolution of voids by tracking each void across consecutive loading cycles, as a consequence of the non-affine displacements of particles nearby. For a time series of binarized 2D images (\autoref{fgr:void_calc}a), we stack them into a 3D spatial-temporal image, in which each void occupies a continuous volume that can be easily identified, eventually yielding a time history of its area $\tilde{A}_v$. 
We also track events involving several voids merging into one, and a single void splitting into several voids. In this case, we calculate the summed area of all contributing voids for an event, also denoted by $\tilde{A}_v$, so that we can quantify the change in void area, $\Delta\tilde{A}_v$, due to a given event. 

\begin{figure}[t!]
\centering
  \includegraphics[width=0.5\textwidth]{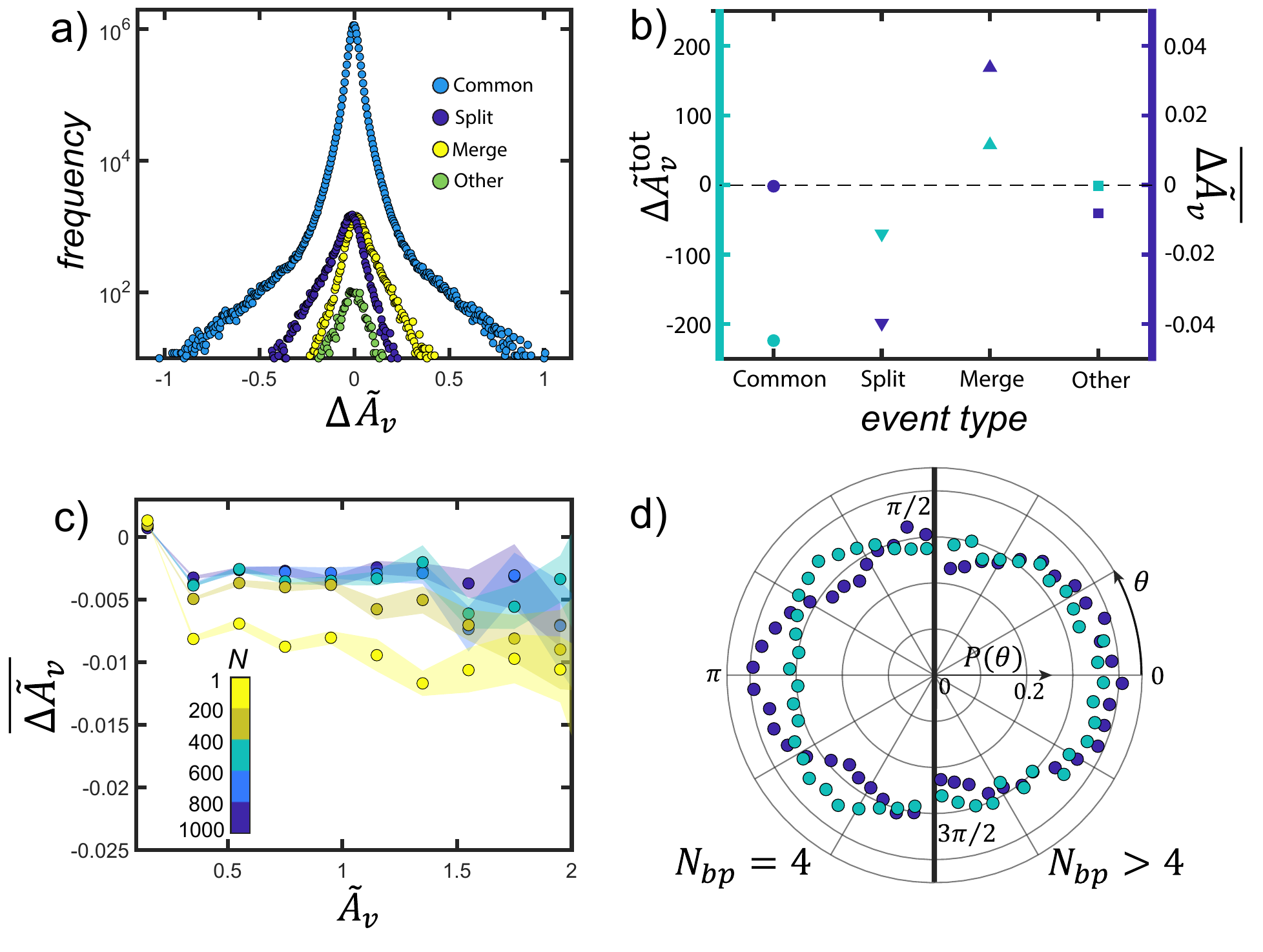}
  \caption{The change of void morphology during cyclic loading. a) Frequency count of the area change for different void behavior types. b) The total void area change (left axis, turquoise) and average individual void area change (right axis, dark blue) due to specific event type. c) The change in void area vs. current void area for Common-type voids. The shaded regions indicate standard error across experiments. d) A histogram of the orientations of the principal axes for voids with $N_\textrm{bp} > 4$ (right half) and $N_\textrm{bp} = 4$ (left half). Data points taken at $N=0$ are shown in dark blue, and data points taken at $N=1000$ are shown in turquoise.
  }
  \label{fgr:void_morpho}
\end{figure}

Following observations of void tendencies, we classify the morphological change of voids into four types of events. 
We use Common-type to denote a change in a void's area as its constituent particles rearrange, but without change to its connectivity in the 3D spatial-temporal image. For this analysis, we do not include voids with $N_\textrm{bp}=3$.

Moving on, the Split-type denotes an event in which a single void splits into multiple voids, while the Merge-type corresponds to multiple voids merging into a single void. Finally, the Other-type represents a complex morphology change, where multiple voids reorganize into multiple new voids, in a way that cannot be decomposed into simple merging or splitting. For all four types, we calculate $\Delta\tilde{A}_v$ using a small loading cycle interval of $\Delta N=2$. 

Under this definition, we sampled events throughout a cyclic loading test and from 14 tests with different initial packing structures. \autoref{fgr:void_morpho}a shows a frequency count of the change in void area $\Delta\tilde{A}_v$ for the four types of events, revealing that Common-type events are the most common by several orders of magnitude. Similar numbers of Merge-type and Split-type events are observed, followed by the least frequent Other-type events.

For a given type of void event,
we then calculate the average area change per void, $\overline{\Delta\tilde{A}}_v$, and the total area change, $\Delta\tilde{A}_v^\mathrm{tot}$, see \autoref{fgr:void_morpho}b.
On average, Split-type and Merge-type events result in negative and positive area change, respectively. Common-type events have a slightly negative $\Delta\tilde{A}_v$ on average and the associated magnitude $|\overline{\Delta\tilde{A}}_v|$ is significantly smaller than that of Merge-type and Split-type events, which are similar to each other. Despite this, Common-type events are so much more frequent that their contribution to the total area change, $\Delta\tilde{A}_v^\mathrm{tot}$, is greater than any other event type. Additionally, as Merge-type and Split-type events occur at similar frequencies and with opposite signs in $\overline{\Delta\tilde{A}}_v$, their contributions to $\Delta\tilde{A}_v^\mathrm{tot}$ nearly cancel out. Finally, although Other-type events have a negative $\overline{\Delta\tilde{A}}_v$, their low frequencies prevent any significant $\Delta\tilde{A}_v^\mathrm{tot}$. 

As Common-type events have the greatest contribution to the densification of the raft, we examine the dependence of their area change, $\Delta \tilde{A}_v$, on their current size $\tilde{A}_v$. We binned all applicable voids by $\tilde{A}_v$, and calculated the bin-averaged $\Delta \tilde{A}_v$ for voids sampled at various $N$, which is shown in \autoref{fgr:void_morpho}c. The results show that for the smallest area bin, which is almost entirely composed of voids with $N_\textrm{bp} = 4$, a small, positive, area change is observed. This indicates that these voids grow in small increments more than they shrink, consistent with the small radially outward displacement fields observed in \autoref{fgr:void_dyn}b. It is worth pointing out that the area change due to a void switching from $N_\textrm{bp} = 4$ to $N_\textrm{bp} = 3$ is not considered Common-type and is therefore not represented in \autoref{fgr:void_morpho}.

For larger voids, however, the change in area is negative, and nearly identical across different values of $\tilde{A}_v$. The radially inward displacement field from \autoref{fgr:void_dyn}c corroborates this observation. The sign switch in $\overline{\Delta\tilde{A}}_v$ between the two smallest area bins provides further evidence of differing void behavior between $N_\textrm{bp} > 4$ ($\tilde{A}_v > 0.29$) voids and $N_\textrm{bp} \leq 4$ ($\tilde{A}_v \leq 0.29$) voids. This aligns with the tendency of $N_\textrm{bp} > 4$ voids to contribute significantly more to the raft's dynamics than $N_\textrm{bp} \leq 4$ voids, which was first observed in \autoref{fgr:void_dyn}e.  

\begin{figure*}[t!]
\centering
  \includegraphics[width=0.95\textwidth]{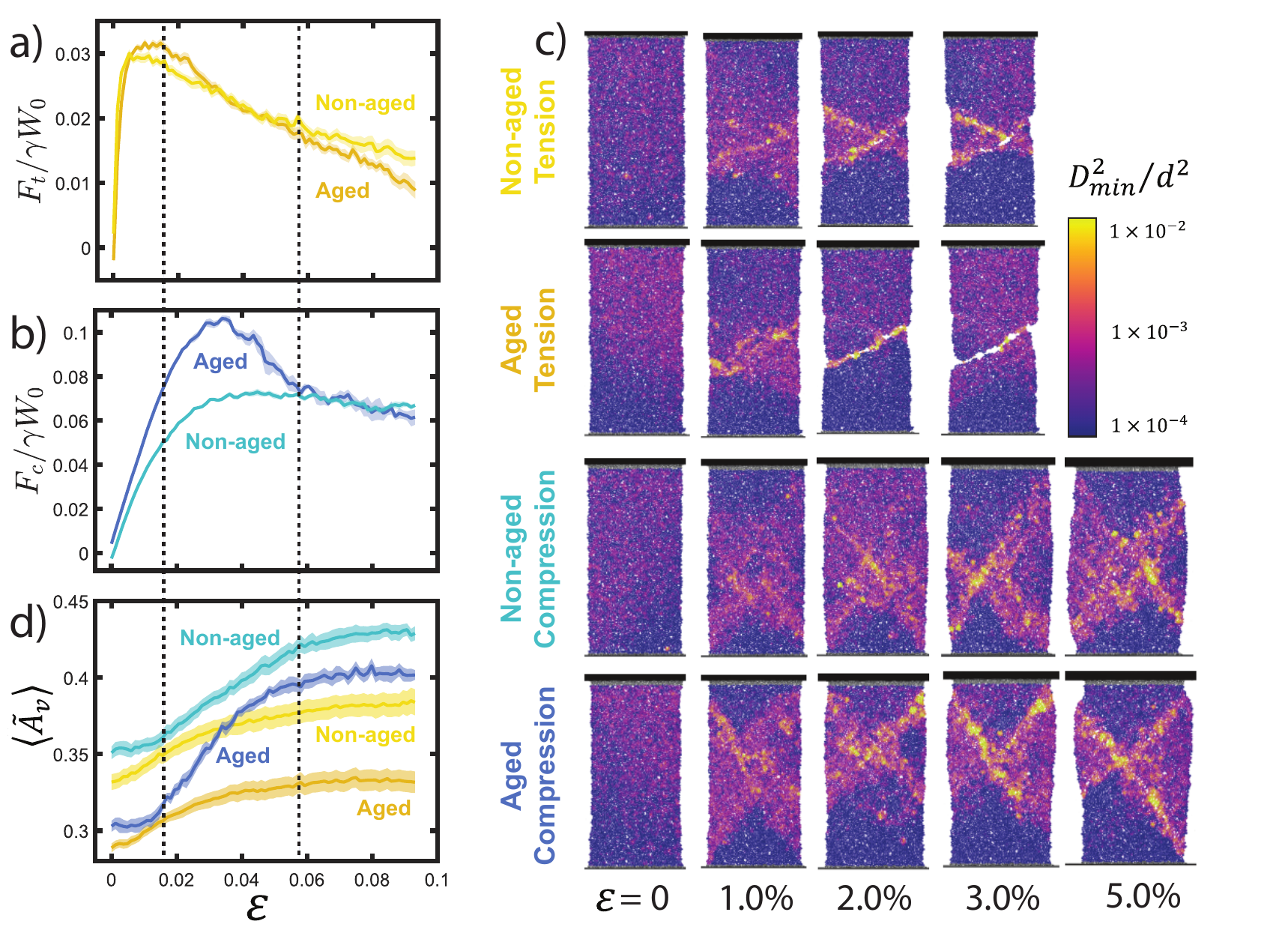}
  \vspace{-2 mm}
  \caption{Uniaxial tension and compression tests on the non-aged and aged rafts. 
  a) Stress-strain curves for the non-aged (light orange) and aged (dark orange) rafts in tensile tests, averaged over 16 and 14 repetitions for each experiment, respectively.
  b) Stress-strain curves for the non-aged (light blue) and aged (dark blue) rafts in compression tests, averaged over five and eight repetitions for each experiment, respectively. 
  %\HX{TO DO: double check that we actually have eight repetitions of aged compression experiment. Also, for numbers smaller than ten, just spell out the word.}
  c) A visualization of the particle $D^2_{\mathrm{min}}$ in the non-aged and aged rafts undergoing compression and tension, shown for various global strains.
  d) The average void area vs. global strain for the cases in a) and b) using the same color scheme. The shaded regions in a), b), and d) indicate test-to-test fluctuations.
  }
  \label{fgr:failure}
\end{figure*}

There is also a dependence of $\overline{\Delta\tilde{A}}_v$ on $N$, showing that $|\Delta \tilde{A}_v|$ is smaller for larger $N$, similar to the dynamic slow-down observed in \autoref{fgr:raft_struc}. This result, along with those from \autoref{fgr:raft_struc}, suggests that most of the structural change occurs in the earlier stages of cyclic loading.

Finally, void orientation can be investigated as a result of aging. To determine the orientation of a single void, the pixels composing a void from an experimental image are used to identify the major axis of an ellipse with the same normalized second central moments. The orientation, $\theta$, is then defined as the angle that this axis makes with another axis running parallel to the fixed boundaries of the raft. We can then separate the orientation of voids into two categories, $N_\textrm{bp} = 4$ and $N_\textrm{bp} > 4$, to distinguish whether the behaviors are different, motivated by the differences between $N_\textrm{bp} = 4$ and $N_\textrm{bp} > 4$ voids observed in \autoref{fgr:void_dyn} and \autoref{fgr:void_morpho}c. We can then separate them further into groups of $N=0$ or $N=1000$ to observe the effects of mechanical aging.

We see that all void sizes initially show a preference toward $\theta= 0$, especially for the case of voids with $N_\textrm{bp} = 4$. At $N=1000$, voids with $N_\textrm{bp} > 4$ tend to be oriented perpendicularly more often than at $N=0$, although a preference for $\theta = 0$ remains. Voids with $N_\textrm{bp} = 4$ at $N=1000$, however, are found most frequently at multiples of $\theta = \frac{\pi}{4}$, coinciding with the direction of maximum shear stress in a uniaxially deformed material, and also similar to the orientation of the shear bands observed in the next section. The histogram showing the probability density of all void orientations is shown in \autoref{fgr:void_morpho}d.

Therefore, voids with $N_\textrm{bp} > 4$ experience larger negative area changes and contribute significantly more to the density change of the raft than voids with $N_\textrm{bp} = 4$. Voids with $N_\textrm{bp} = 4$, however, do not remain static, instead exhibiting a dramatic shift in their orientations over the course of aging, in contrast to the minor orientation change observed for voids with $N_\textrm{bp} > 4$.

\section{Implications of aging on raft failure}
\label{Sec7}

To investigate how the structural changes induced by cyclic loading affect the yielding behavior of the raft, we subject aged rafts (loaded to $N=1000$), as well as non-aged rafts (without any cyclic loading history), to uniaxial tension and uniaxial compression. We measure $F_{i}/\gamma W_0$ against the global strain $\epsilon=\delta_{i}/L_0$, where $i=t$ for tension and $i=c$ for compression. We also visualize the deformation behavior of the rafts by overlaying experimental images with particles colored by their respective $D^2_{\mathrm{min}}$ values, which are calculated over a strain interval $\Delta\epsilon=0.001$. The results are displayed in \autoref{fgr:failure}.

In tensile tests, both aged and non-aged rafts display a similar tensile stiffness, measured as the slope of the initial increment of the loading curve, until the onset of yielding at $\epsilon\approx0.01$. However, the strength of the aged rafts, defined as the maximum value of $F_{t}/\gamma W_0$, is slightly larger than that of the non-aged rafts, which is shown in \autoref{fgr:failure}a. Also associated with the aged rafts is a faster decay in the tensile stress following $\epsilon>0.01$. These results indicate that aged rafts are slightly stronger and more brittle during tensile deformation. The visualization of two example rafts (\autoref{fgr:failure}c), representing aged and non-aged rafts, reveals shear banding that is narrower in the aged case through more concentrated regions of high $D^2_{\mathrm{min}}$.  

While the loading curves were similar up to yielding in the tensile tests of the aged and non-aged rafts, there are clear differences between the two types of compression tests as shown in \autoref{fgr:failure}b. The relation between $F_{c}/\gamma W_0$ and $\epsilon$ indicates a significantly higher stiffness for the aged rafts, as well as a significantly higher yield strength at $\epsilon\approx0.03$, which coincides with a unified yield strain value of a variety of disordered solids.\cite{cubuk2017structure} 
Beyond the yielding point, the two loading curves are rather distinct. While the stress for the non-aged rafts continuously increases until it plateaus, a significant overshoot exists in the curve for the aged raft, followed by a stress drop toward the same plateau as in the non-aged case, indicating a history-independent steady state-like behavior.
The stress drop in the aged rafts before the plateau is indicative of a brittle behavior, while the continuous stress increase in the non-aged rafts indicates ductile behavior. 
Correspondingly, the shear bands tend to be narrower in the aged rafts, with the particles inside having higher $D^2_{\mathrm{min}}$, as seen in \autoref{fgr:failure}c.

Another difference between tension and compression tests manifests in the orientation of the shear band during failure. The orientation of the shear band during compression is above $\pi/4$, and below $\pi/4$ during tensile tests, suggesting that volumetric effects and friction may be important for the yielding process.\cite{bardet1990comprehensive,karimi2018correlation}

As a final analysis, we investigate the behavior of voids during the uniaxial loading tests by examining the average void area $\langle\tilde{A}_v\rangle$ vs $\epsilon$, which can be calculated using \autoref{eq:void} and normalized in the same way as in \autoref{sec:five}.
In all tests, $\langle\tilde{A}_v\rangle$ increases with $\epsilon$, showing that the raft generally dilates regardless of the deformation mode. The plot of $\langle\tilde{A}_v\rangle$ against $\epsilon$ is shown in \autoref{fgr:failure}d.
For tension, this is a straightforward outcome due to the imposed extensional volumetric strain. 
%It is expected that the tension tests will have the lowest amount of dilation, as regions outside the shear band remain aged, even during failure\cite{parmar2019strain}
%, and due to the positive volumetric strain component, free volume is being added to the system, reducing the need for dilation. 
The dilation observed in the compression tests, however, indicate that the deviatoric part of the imposed uniaxial deformation dominates the volumetric part, as deviatoric deformation often induces dilation in dense granular materials. This suggests that the densification of the raft played a role in its lowered ductility, as greater dilation was required to reach steady-state behavior, which is a signature of brittle-like failure.

\section{Conclusions}
\label{sec:cls}

We have demonstrated that a cohesive granular raft subject to small amplitude cyclic loading experiences structural evolution analogous to aging observed in thermal systems, caused in part by a long-term logarithmic increase in the raft's density. This density change is only possible due to the presence of free boundaries and uniaxial loading conditions, which promotes changes to volume. Other parameters tied closely to the structure of the raft, such as the bond orientation parameter and structural heterogeneity, also change logarithmically with cycle number, consistent with the dynamic slow-down in the cage-relative $MSD/d^2$ and $\langle D^2_{min}\rangle/d^2$.

To address the source of structural evolution during aging, we monitor voids, defined as discrete areas associated with gaps in the particle packing structure. We see that the area of voids also decreases logarithmically, but that large and small voids exhibit very different behaviors. Large voids tend to displace nearby particles toward their center during aging, shrinking the void's area and increasing the density of the raft. Additionally, using $\bar{D}^2_{min,L}$ as a metric for structural evolution, we see that larger void areas tend to induce more structural change in their close vicinity. Further analysis suggests that although large scale rearrangements individually contribute the most to the area change of a void, it is the smaller rearrangements that dominate the density change of the raft. Thus, the structural change of the raft is associated closely with the shrinking of large voids, and the density change is most influenced by small, but commonplace, rearrangements to those voids. 

Smaller voids, which can be classified as having four constituent boundary particles, behave very differently. On average, these voids tend to grow slightly, but on a magnitude appreciably smaller than the area change associated with larger voids. Despite this small contribution to total raft density, we see that these voids are not inactive, as a distribution of their orientations changes to align primarily with an angle of $\pi/4$ radians with the horizontal after 1000 loading cycles, despite initially aligning with the horizontal axis. Larger voids, meanwhile, see very little change to the distribution of their orientations, suggesting that small voids are more susceptible to changes in orientation.   

The structural changes to the raft culminate in altered deformation properties and failure mechanisms. Aged rafts undergoing tensile tests have a higher strength than non-aged rafts, but also exhibit more brittle behavior. A similar trend is observed under compression, with the added observation of increased stiffness and a stress overshoot, indicating brittle failure and a dilating material. This additional dilation is due to the increased density of the raft, and thus the shrinking of large voids. Regardless of preparation history, however, all rafts under compression tend to a steady state behavior for large strains. Investigating the area of voids during these uniaxial tests confirms particularly strong dilation during the compression of aged rafts. Also observed is a shear band close to $\pi/4$ radians for all failure tests, but with a higher or lower angle depending on whether the test was under compression or tension, respectively. 

Our experiment has potential applications in the study of yielding disordered solids, as we can intentionally trigger a stress overshoot during yielding via well-controlled cyclic loading. Our methods for monitoring the raft's microscopic structure and dynamics could be used to better understand how the preparation history of an out-of-equilibrium disordered solid influence the nature of its yielding transition.\cite{ozawa2018random,moorcroft2011age,berthier2025yielding} The capillary-induced particle attraction and the particle friction also renders our experimental system suitable to study whether complex particle interactions give rise to rugged energy landscapes of an amorphous solid, exhibiting possibly perpetual irreversibility.\cite{xiao2022probing,yuan2024creep,li2015deformation}

\section*{Conflicts of interest}
There are no conflicts to declare.

\section*{Acknowledgements}
The authors would like to acknowledge the funding from the National Science Foundation grant MRSEC/DMR-1720530 and grant CMMI-2519512. We also
thank Andrea J. Liu for helpful discussions.
%\nocite{*}

\bibliography{rsc}% Produces the bibliography via BibTeX.

\end{document}